\documentclass[
aps,%
12pt,%
final,%
oneside,%
onecolumn,%
nobibnotes,%
nofootinbib,%
superscriptaddress,%
centertags,
floatfix,
secnumarabic]%
{revtex4}%
\usepackage{graphicx}
\usepackage{bm}
\usepackage{slashed}
\usepackage{subfigure}
\usepackage{multirow}
\RequirePackage[english]{babel}
\usepackage{amsmath}
\usepackage{cleveref}

\usepackage[normalem]{ulem}
\usepackage{xcolor}

\begin{document}

\selectlanguage{english}
\normalsize

\title{One-loop corrections to the processes $e^+e^- \to \gamma, Z\to  J/\psi\: \eta_c$ and $e^+e^- \to Z \to  J/\psi\: J/\psi$}
\author{\firstname{A.~V.}~\surname{Berezhnoy}}
\email{Alexander.Berezhnoy@cern.ch}
\affiliation{SINP MSU, Moscow, Russia}

\author{\firstname{I.~N.}~\surname{Belov}}
\email{ilia.belov@cern.ch}
\affiliation{SINP MSU, Moscow, Russia}
\affiliation{Physics department of MSU, Moscow, Russia}

\author{\firstname{S.~V.}~\surname{Poslavsky}}
\email{Stanislav.Poslavskii@cern.ch}
\affiliation{NRC “Kurchatov Institute”–IHEP, Protvino, Russia}

\author{\firstname{A.~K.}~\surname{Likhoded}}
\email{Anatolii.Likhoded@ihep.ru}
\affiliation{NRC “Kurchatov Institute”–IHEP, Protvino, Russia}

\begin{abstract}
\small
The cross sections of  $J/\psi\: \eta_c$ and $J/\psi\: J/\psi$ production  in $e^+e^-$ annihilation are calculated within a framework of a one-loop approximation near  $Z$-boson pole and at higher energies as well. Both intermediate bosons, $\gamma$ and $Z$, are included.  It is found that at  $Z$ mass the next-to-leading contribution increases the production cross-sections by a factor of 3.5~.
\end{abstract}

\maketitle

\section{INTRODUCTION}

An associative production of $J/\psi\: \eta_c$ in  $e^+e^-$ annihilation was studied experimentally in details by Belle and BaBar Collaborations at the energy around 10.6 GeV~\cite{Abe:2004ww,Aubert:2005tj}. It turned out that the cross-section value predicted within the LO order approximation was by a factor of more than 5 smaller than the experimental measurement. This significant difference has initiated an intensive study of various corrections to the LO mechanism.

Two sources of corrections were mainly concerned. The first one is the internal motion of the quarks inside quarkonium (see the Ref.~\cite{Bondar:2004sv}, which initiated the discussion on this contribution and then been followed by Refs.~\cite{Braguta:2005kr,Berezhnoy:2006mz,Braguta:2006nf,Bodwin:2006dm,Ebert:2006xq,Berezhnoy:2008zz,Ebert:2008kj,Braguta:2008hs,Braguta:2008tg,Sun:2009zk,Braguta:2012zza, Sun:2018rgx}). The second source of enhancement is due to QCD loop corrections and the one-loop corrections have already been determined~\cite{Zhang:2005cha,Gong:2007db}. Nowadays, the corrections are known up to two-loops accuracy at the energy of $B$-factories~\cite{Feng:2019zmt}. There are results, where both types of corrections were considered and applied~\cite{Dong:2012xx, Xi-Huai:2014iaa}. Evaluating the theoretical results in this field, one tends to believe that indeed both mechanisms are required to correctly describe the data.

Future charmonia studies are in plans of two big projects, International Linear Collider (ILC) and Future Circular Collider (FCC). Both projects will make available $e^+e^-$ collisions at $Z$ mass and above, specifically the energy range announced for FCC extends from $\sqrt{s} = 90~\text{GeV}$ to $400~\text{GeV}$, while the collision energy of ILC has to be tuned to $250~\text{GeV}$. Furthermore, the studies of $Z$-boson decays into two charmonia are certainly of an interest for the running LHC experiments. The $J/\psi \: \eta_c$ and  $J/\psi\:  J/\psi$ pair production near $Z$-boson pole  were calculated in LO approximation~\cite{Hagiwara:2003cw, Chen:2013mjb}. Moreover,  $Z$-boson decay mode to two charmonia was studied in the framework of a light cone formalism, what allowed to include the internal motion of quarks inside charmonium~\cite{Likhoded:2017jmx}. Complementing the aforementioned results, another type of the one-loop correction is introduced in this report, specifically the $J/\psi \: \eta_c$ and  $J/\psi\: J/\psi$ pair production in the $e^+e^-$ annihilation is computed up to the  one-loop accuracy including both $Z$ boson and photon, 
$$\notag
\begin{cases}
e^+e^- &\xrightarrow{\gamma^*,Z^*}\ {J\psi \: \eta_c}\,, \\
e^+e^- &\xrightarrow{Z^*}\ {J/\psi\: J/\psi}\,. \\
\end{cases}
$$

The theoretical study on $J/\psi\: J/\psi$ pair production at $e^+e^-$ collisions via a double photon exchange has been performed~\cite{Bodwin:2002fk,Gong:2008ce}. Unfortunately, the attempts to make the calculations of the process at $B$-factories had no progress~\cite{Abe:2003ja,Abe:2004ww, Aubert:2005tj}. In the present report, the $e^+e^-$ annihilation via a single boson only is concerned.

The report follows up the previous research on $B_c$-pair production in $e^+e^-$ annihilation~\cite{Berezhnoy:2016etd}. Meanwhile, the $B_c$-pair production via $\gamma \gamma$-fusion  has also been covered in Ref.~\cite{Chen:2020dtu}.

\section{THE METHOD}

The discussed  production of the charmonium pair via a single boson exchange is affected by  several selection rules. First of all, neither photon nor $Z$ may decay to two identical $\eta_c$ mesons, since the $\eta_c$ pair must be in a P-wave state with a symmetric wave function, what is impossible.  Second, $J/\psi\: J/\psi$ pair can not be produced by a single photon exchange due to a charge parity conservation. Similarly to the photon case, the vector part of the $Z$-boson vertex does not contribute to the $J/\psi\: J/\psi$ production. Third, due to the charge parity conservation the axial part of the $Z$-boson vertex does not contribute to the $J/\psi\: \eta_c$ amplitude. The listed above selection rules are explicitly reproduced in the calculations presented below, hence providing the additional verification of the procedure.

The production of  double heavy bound states is effectively described by the NRQCD factorization~\cite{Bodwin:1994jh}. The factorization formalism is introduced to factor out the perturbative degrees of freedom, therefore to separate the production mechanism into hard (short distance) and soft (long distance) subprocesses. Given the fact that $m_c >> m_c v$, where $v$ is the velocity of $c$-quark in charmonium, the short distance interaction corresponds to the perturbative part  of $c\bar c$-pair production, whereas the long distance interaction describes the bound state formation and dynamics.

In our computations of the $J/\psi\: \eta_c$ production matrix elements, we start from the matrix element of four heavy quark production $e^+e^-\to c(p_c) \bar c(p_{\bar c}) c(q_c) \bar c(q_{\bar c})$ with heavy quarks and antiquarks defined on their mass shells: $p_c^2=p_{\bar c}^2=q_c^2=q_{\bar c}^2=m_c^2$. As we assign $v=0$ before the projection  onto the bound states, the momentum $P$ of the vector charmonium and the momentum $Q$ of the pseudoscalar charmonium  are related with the heavy quark momenta as follows below, 
\begin{align}\label{momenta}
&J/\psi  \begin{cases} 
&p_c = P/2 \\ &p_{\bar c} = P/2 \\
\end{cases}
&\eta_c\ \begin{cases}
&q_c = Q/2 \\ &q_{\bar c} = Q/2\,.
\end{cases}
\end{align}

To construct the bound states, we replace the spinor products $v(p_{\bar c})\bar u(p_c)$ and $v(q_{\bar c})\bar u(q_c)$ by the appropriate covariant projectors for color-singlet, spin-singlet and spin-triplet states as per
\begin{align}
\label{proj}
     &\Pi_{J/\psi}(P,m)=\frac{\slashed P- m}{2\sqrt{m}}\ \slashed \varepsilon \otimes \frac{\boldsymbol 1}{\sqrt{3}}\,,     &\Pi_{\eta_{c}}(Q,m)=\frac{\slashed Q- m}{2\sqrt{m}}\gamma^{5}\otimes \frac{\boldsymbol 1}{\sqrt{3}}\,, 
 \end{align}
where $m=2m_c$, $\varepsilon$ is the polarization of the  $J/\psi$ meson, satisfying the following constraints: $\varepsilon\cdot \varepsilon^*=-1$, $\varepsilon\cdot P=0$.

In exactly the same way, we express the matrix element of the two vector charmonia, denoting their momenta by $P_1$ and $P_2$, and their polarizations by $\varepsilon^1$ and $\varepsilon^2$.

\begin{figure}[ht]  
 \centering 
 \subfigure[]{
\includegraphics[width=0.35\linewidth]{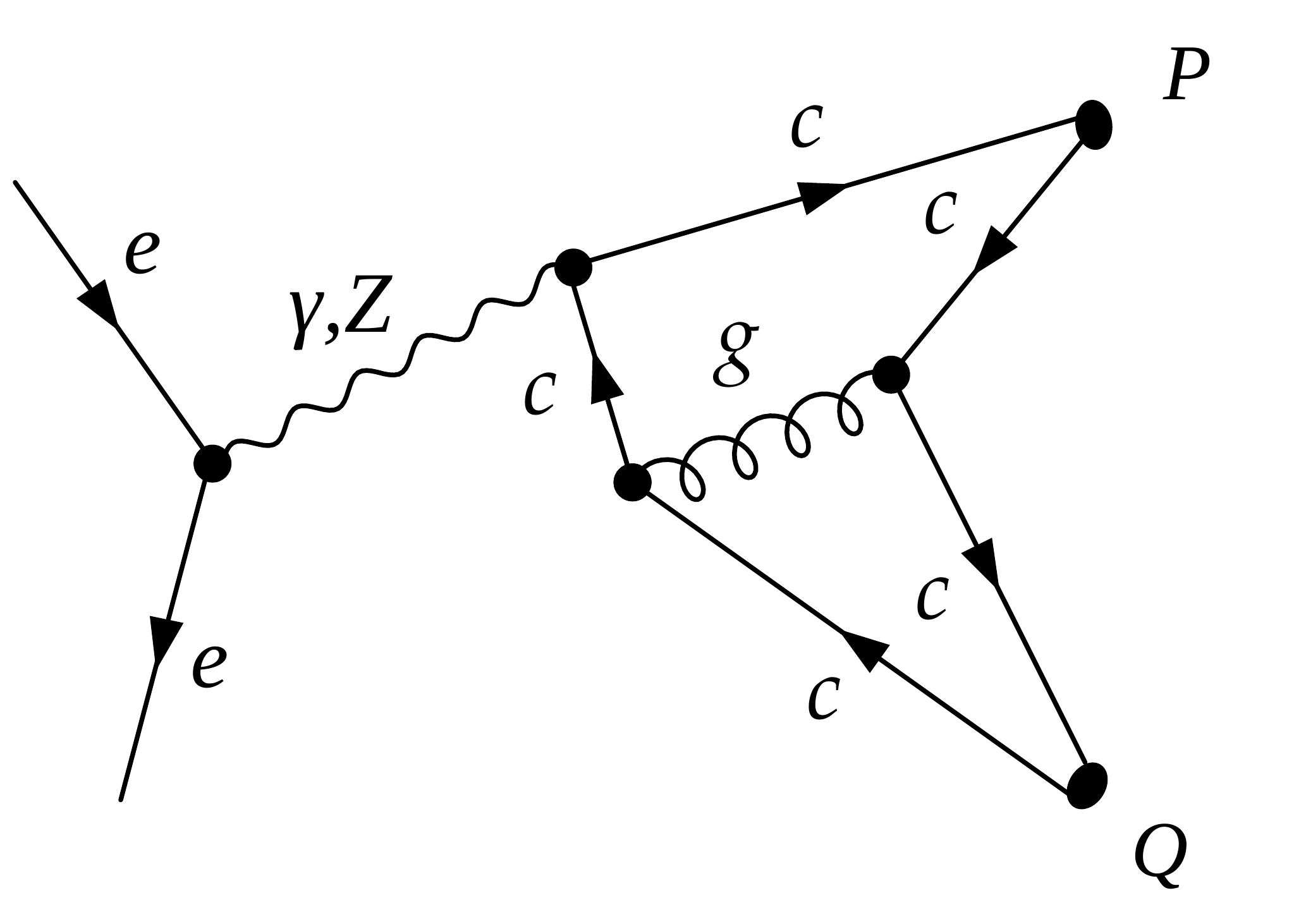} \label{fig:picLO}}  
\hfill
\subfigure[]{
\includegraphics[width=0.3\linewidth]{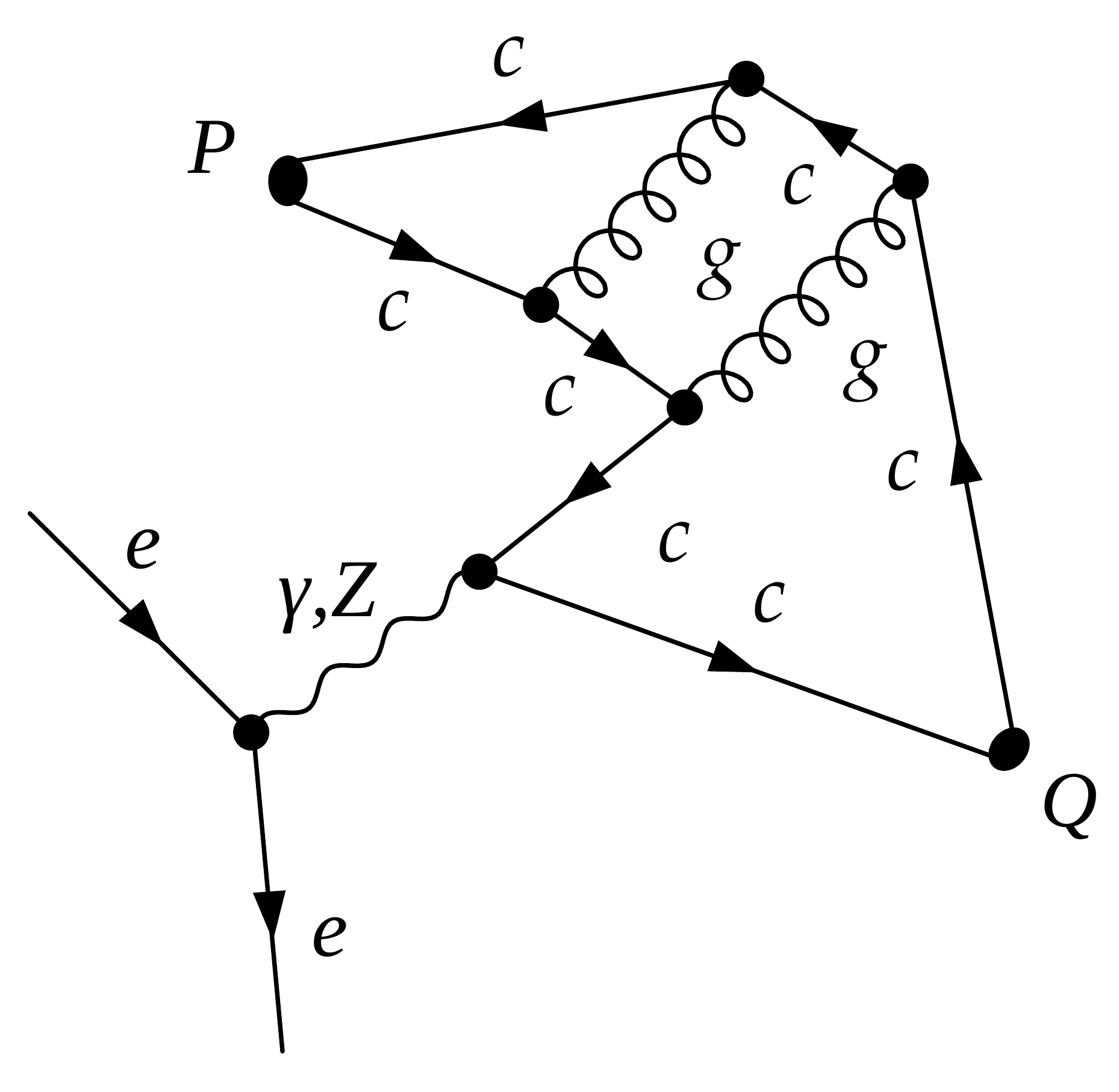} \label{fig:picNLO1}} 
\hfill
\subfigure[]{
\includegraphics[width=0.28\linewidth]{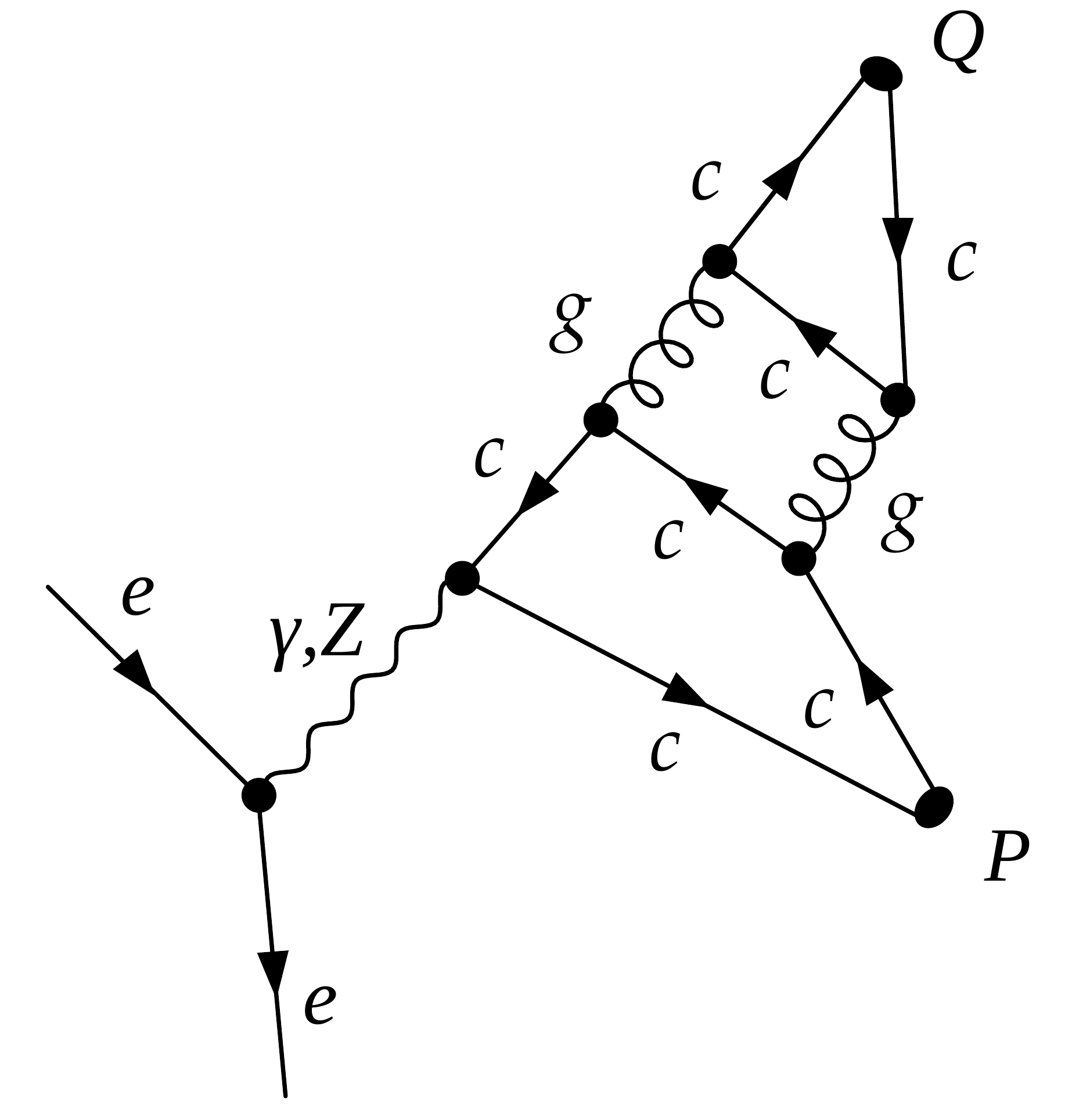} \label{fig:picNLO2}}
\caption{The examples of diagrams for $e^+e^-\to\ J/\psi\: \eta_c$ process with one and two traces: the LO diagram \subref{fig:picLO};  the NLO diagrams \subref{fig:picNLO1}~and~\subref{fig:picNLO2}.} 
\label{fig:projectwaynew}
\end{figure}

The operators (\ref{proj})  close the fermion lines into traces. The contributions of LO diagrams to the amplitude always contain only one trace, while the NLO  contributions contain one or two traces as illustrated at Figure~\ref{fig:projectwaynew}.

The factorized matrix elements have the forms specified below in Eqs.~\ref{simple} and ~\ref{simple1}, 
\begin{align}
{\cal A}\left[e^+e^- \to J/\psi(P)\;\eta_c(Q)\right] &= \frac{1}{4\pi }R_{J/\psi}(0) R_{\eta_c}(0) \cdot \bigl.{\cal M}^{\mu}\left(P,Q\right)\varepsilon_{\mu}\label{simple}\,,  \\
{\cal A}\left[e^+e^- \to J/\psi(P_1)\; J/\psi (P_2)\right] &= \frac{1}{4\pi }R_{J/\psi}^2(0) \cdot \bigl. {\cal M}^{\mu\nu}\left(P_1,P_2\right)\varepsilon_{\mu}^1\varepsilon_{\nu}^2\label{simple1}\,, 
\end{align}
where ${\cal M}^{\mu}\left(P,Q\right)\varepsilon_{\mu}$ and  ${\cal M}^{\mu\nu}\left(P_1,P_2\right)\varepsilon_{\mu}^1\varepsilon_{\nu}^2$
are the hard production matrix elements of the two quark-antiquark pairs, projected on the quark-antiquark states with zero relative velocities and the appropriate
quantum numbers using projectors (\ref{proj}); 
and  $R_{J/\psi}(0)$, $R_{\eta_{c}}(0)$  are the radial wave function values at  origin.

The added caveat is that a real gluon radiation does not contribute to the NLO corrections in the studied processes, since within the applied approximation both heavy quark pairs are produced in the color singlet states.
Thereby the  NLO QCD corrections include  only the  contribution of the interference between the LO amplitudes and the one loop amplitudes. The total squared amplitude is of the order of ${\cal O}(\alpha^2\alpha_S^3)$. It contains the following seven terms, specifically
\begin{multline}\label{interference}
|{\cal A}|^2 = |{\cal A}_{\gamma}^{LO}|^2 + |{\cal A}_Z^{LO}|^2 + 2Re\left({\cal A}_{\gamma}^{LO}{\cal A}_Z^{LO*}\right) +\\ 
 + 2Re\left({\cal A}_{\gamma}^{LO}{\cal A}_{\gamma}^{NLO*}\right) + 2Re\left({\cal A}_Z^{LO}{\cal A}_Z^{NLO*}\right) + 2Re\left({\cal A}_{\gamma}^{LO}{\cal A}_Z^{NLO*}\right) + 2Re\left({\cal A}_Z^{LO}{\cal A}_{\gamma}^{NLO*}\right).
\end{multline}

The so-called ``on shell'' scheme is used for renormalization of masses and spinors and $\overline{MS}$ scheme is adopted for coupling constant renormalization as per
\begin{align}\label{mrenorm}
Z_m^{OS} &= 1 - \frac{\alpha_s}{4\pi}C_F C_{\epsilon}\left[\frac{3}{\epsilon_{UV}} + 4\right] + {\cal O}(\alpha_s^2)\,, \\ \label{wfrenorm}
Z_2^{OS} &= 1 - \frac{\alpha_s}{4\pi}C_F C_{\epsilon}\left[\frac{1}{\epsilon_{UV}} + \frac{2}{\epsilon_{IR}} + 4\right] + {\cal O}(\alpha_s^2)\,, \\ \label{grenorm}
Z_g^{\overline{MS}} &= 1 - \frac{\beta_0}{2}\frac{\alpha_s}{4\pi}\left[\frac{1}{\epsilon_{UV}} -\gamma_E + \ln(4\pi)\right] + {\cal O}(\alpha_s^2)\,,
\end{align}
where $C_{\epsilon} = \left(\frac{4\pi\mu^2}{m^2}e^{-\gamma_E}\right)^{\epsilon}$ and  $\gamma_E$ is the Euler constant.

The  counter-terms are obtained from the leading order diagrams as follows below,
\begin{equation}
{\cal A}^{CT} = Z_2^2{\cal A}^{LO} \Biggr|_{\substack{\boldsymbol{ m \to Z_m m} \\\boldsymbol{g_s \to Z_g g_s}}}.
\end{equation}

The  NLO amplitude ${\cal \tilde A}^{NLO}$ has been calculated using the physical spinors and masses, as well as the physical value of coupling constant. The isolated singularities are further canceled with the singular parts of ${\cal A}^{CT}$ so that ${\cal A}^{NLO}={\cal \tilde A}^{NLO}+{\cal A}^{CT}$ remains finite for the renormalized amplitude.

\section{CALCULATION DETAILS}
For technical reasons, we calculate separately the amplitudes of the $e^+e^-$ fusion into the virtual $Z$ boson and photon, and consequently the amplitudes of the $Z$-boson and photon transitions into charmonia.

\begin{figure}[h!]
 \centering
\includegraphics[width = 0.97\linewidth]{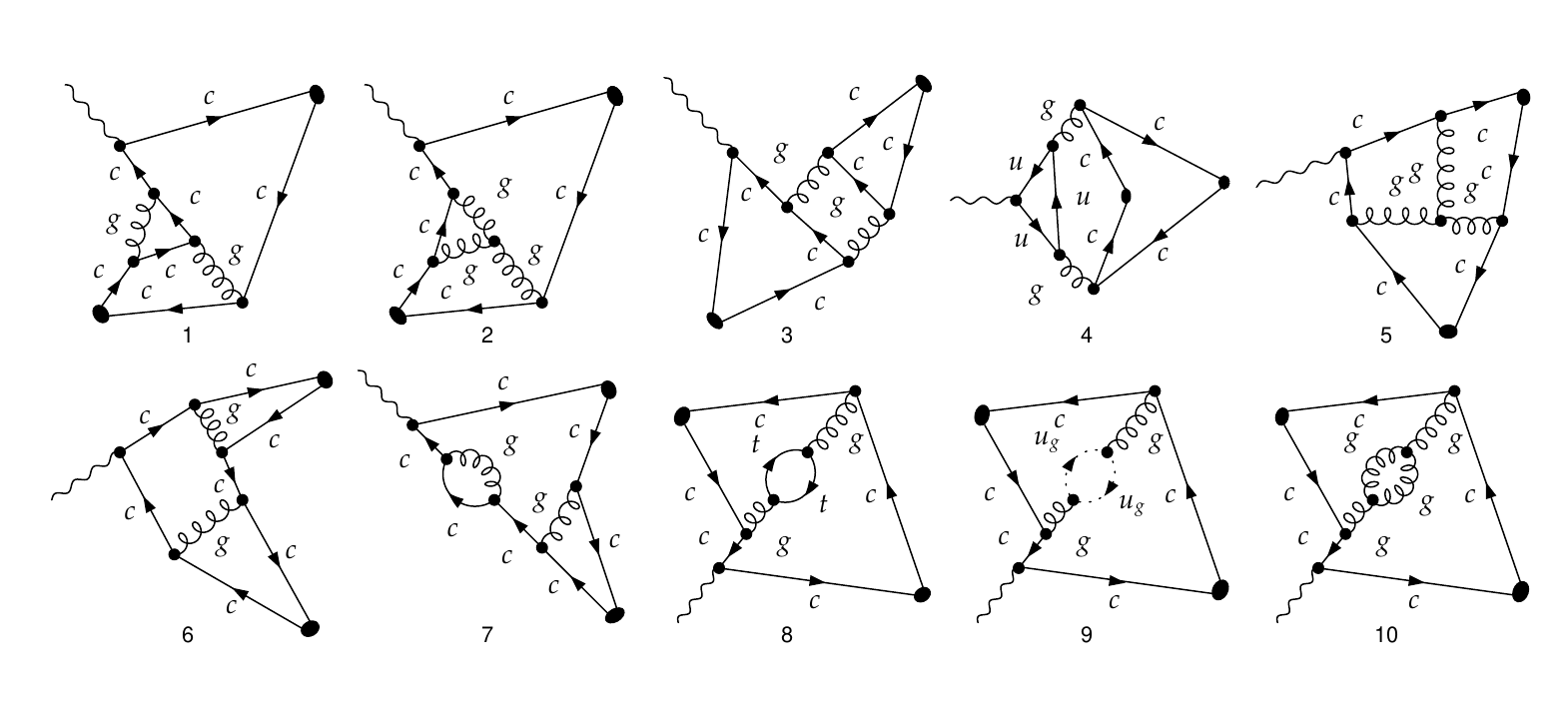}
\caption{ The examples of NLO diagrams for the processes $\gamma^*,Z^*\to J/\psi\: \eta_c\left(J/\psi\: J/\psi\right)$.}
\label{fig:diags}
\end{figure}

There are 4 LO diagrams and 86 one loop diagrams  for the $Z^*$ decay and the same number of diagrams for the $\gamma^*$ decay according to Figure~\ref{fig:diags}. The  diagrams and the corresponding analytic expressions are generated with the  \texttt{FeynArts}-package~\cite{Hahn:2000kx} in Wolfram Mathematica.

The following toolchain is used  for the computations: \texttt{FeynArts} $\to$ \texttt{FeynCalc}~\cite{Shtabovenko:2020gxv} (\texttt{FeynCalcFormLink}~\cite{Feng:2012tk}, \texttt{TIDL}) $\to$ \texttt{Apart}~\cite{Feng:2012iq} $\to$ \texttt{FIRE}~\cite{Smirnov:2008iw} $\to$ \texttt{X}-package~\cite{Patel:2016fam}. The amplitudes  generated with \texttt{FeynArts} are further processed with \texttt{FeynCalc} providing algebraic computations with Dirac and color matrices. The traces are computed with  \texttt{FeynCalc} and \texttt{FORM}. The \texttt{FORM}  is called from the Wolfram Mathematica within the \texttt{FeynCalFormLink} interface.

The Passarino-Veltman reduction is performed using the \texttt{TIDL} library, which is a part of \texttt{FeynCalc}.  After this procedure the amplitudes  do not contain the loop momentum $k$ with open Lorentz indices, whereas the amplitudes do contain this  momentum in scalar products only.  The \texttt{\$Apart} function does the extra simplification of the amplitudes by partial fractioning of IR-divergent integrals. Finally, the \texttt{FIRE} package provides the complete IBP reduction of the amplitudes to master integrals, using the strategy mostly based on the Laporta algorithm~\cite{Laporta:2001dd}. The master integrals are then evaluated by substitution of their analytical expressions using the  \texttt{X}-package. The computations are performed analytically, and the numerical values of the parameters are substituted only at the last step.

The conventional dimensional regularization (CDR) scheme with $D$-dimensional momenta  (loop and external) and Dirac matrices are adopted for the calculations. The so-called naive interpretation of $\gamma^5$ is applied: $\gamma^5$ matrices anticommute with all other matrices and therefore, they are  cancelled out in traces with an even number of $\gamma^5$. In traces with an odd number of $\gamma^5$ matrices the remaining $\gamma^5$ are moved to the right and replaced per Eq.~\ref{g5}, 
\begin{equation}\label{g5}
\gamma^5 = \frac{-i}{24}\varepsilon_{\alpha\beta\sigma\rho}\gamma^{\alpha}\gamma^{\beta}\gamma^{\sigma}\gamma^{\rho}\,,
\end{equation}
where $\varepsilon_{\alpha\beta\sigma\rho}$ is either 4- or $D$-dimensional Levi-Civita tensor. It is checked, that results of calculations do not depend on the dimension of Levi-Civita tensor in~\eqref{g5}. The choice of its  dimension has slightly affected the traces evaluation process, but it has no effect on the renormalized amplitudes.

It should be noticed that the diagrams with 
triangle quark loops, \textit{e.g.} the diagram 4 in Figure~\ref{fig:diags}, do not contribute to the $J/\psi\: \eta_c$ production due to the $C$-parity conservation, and we can directly verify this condition in our calculations.

The diagrams with two heavy quark traces, \textit{e.g.} the diagram 3 in  Figure~\ref{fig:diags}, add about 3\% to the total cross section of $J/\psi\: \eta_c$ production and do not contribute to the $J/\psi\ J/\psi$  production cross section at all. 

As explained earlier, the $\eta_c\: \eta_c$ pair can not be produced in the photon or $Z$ decays, indeed this selection rule is directly reproduced in our calculations at both LO and NLO levels.  

After the \texttt{FIRE} reduction  only one-, two- and three-point integrals of types $\boldsymbol{A}_0$, $\boldsymbol{B}_0$, and $\boldsymbol{C}_0$ are left in the amplitudes. Some integrals of types $\boldsymbol{A}_0$ and  $\boldsymbol{B}_0$ do contribute to the amplitude with the singular coefficient $~\frac{1}{D-4}$. Therefore, in general case, one should carefully evaluate these specific integrals: terms proportional to ${\cal O}(\varepsilon)$ in the master integral expansion may contribute to the finite part of the amplitude. However in the considered processes the $\sim\frac{1}{D-4}$ terms cancel each other contrary to the case of $B_c$ pair production~\cite{Berezhnoy:2016etd}. 
The infinite amplitude parts coming from the divergent master integrals $\boldsymbol{A}_0$ and $\boldsymbol{B}_0$ contain only ${\cal O}(1/\varepsilon)$ poles. Working with automatic tools we do not distinguish $\varepsilon_{IR}~\text{and}~\varepsilon_{UV}$: $\varepsilon_{IR}=\varepsilon_{UV}=\varepsilon$.

\begin{table}[ht]
\centering
 \caption{The parameter values.}
\label{tab:parameters}
\begin{tabular}{ccccccc}
\hline
$m_c$ = 1.5~GeV & & $m_b$ = 4.5~GeV & & $m_t$ = 172.8~GeV & & $M_{Z}$ = 91.2~GeV\\
\hline
$R_{J/\psi}^2$ = 1.1~GeV$^3$ & & $R_{\eta_c}^2$ = 1.1~GeV$^3$ & & $\Gamma_Z$ = 2.5~GeV & & $\sin^2\theta_w$ = 0.23 \\
\hline
\end{tabular}
\end{table}

In the presented calculations we use the strong coupling constant evaluated  with two loops accuracy as per
\begin{equation}\label{alpha}
\alpha_S\left(Q\right) = \frac{4\pi}{\beta_0 L}\left(1-\frac{\beta_1\ln L}{\beta_0^2 L}\right)\,,    \nonumber
\end{equation} 
where $L=\ln\left(Q^2/\Lambda^2\right)$, $\beta_0 = 11-\frac{2}{3}N_f$, $\beta_1 = 102 - \frac{38}{3}N_f$; the reference value $\alpha_S\left(M_Z\right) = 0.1179$. $N_f=5$ is chosen for the interaction energies above 30 GeV, and $N_f=4$ is chosen for the interaction energies below 30 GeV.


The same value for the renormalization scale and for the coupling scale is used,  $Q=\mu_R=\mu$.
Calculating the loop amplitudes we assume that  $u$-, $d$- and $s$-quarks are massless. The fine structure constant is fixed in the  Thomson limit $\alpha = 1/137$. The numerical values of other parameters  are oulined in Table~\ref{tab:parameters}.

\section{RESULTS}

Since the analytical expressions for NLO cross sections are too cumbersome, we avoid showing ones in the text while presenting only the numerical values of the cross sections at several energies. The results are encapsulated in the Table~\ref{tab:check_values} below.

On the contrary, the analytical expressions for leading order cross sections are quite simple and we present them below, highlighting the contributions of the $\gamma^*$ decay, the $Z^*$ decay and the interference between $\gamma^*~\text{and}~Z^*$ as per Eqs.~\ref{sig_lo}--\ref{sig_lo_jpsi}, 
\begin{equation}
\sigma_{J/\psi\: \eta_c} = \left.\right.\frac{131072\pi\alpha^2\alpha_S^2 R_{J/\psi}^2R_{\eta_c}^2 (1-4m^2/s)^{3/2}\left(1 + a_{\gamma Z} + a_{Z}\right)}{243\ s^4},
\label{sig_lo} 
\end{equation}
\begin{equation}
\sigma_{J/\psi\: J/\psi} = \frac{32 \pi  \alpha ^2 \alpha_S ^2 R_{J/\psi}^4 (1-4m^2/s)^{5/2} \left(\csc ^4\theta_w-4 \csc^2\theta_w+8\right) \sec^4\theta_w}{27\ s^2 \left(\left(M_{Z}^{2} - s\right)^2+\Gamma ^2 M_{Z}^{2}\right)},
\label{sig_lo_jpsi}
\end{equation}
where
\begin{align}
a_{\gamma Z} &= \frac{\tan^2\theta_w \left(3\csc^4\theta_w-20\csc^2\theta _w+32\right)}{16}\ \frac{s\left(s-M_{Z}^{2}\right)}{\left(M_{Z}^{2}-s\right)^2 + \Gamma^2M_{Z}^{2}},
\label{coeff_gamma_Z}\\
a_{Z} &= \frac{\tan^4\theta_w\left(\csc^4\theta_w - 4\csc^2\theta_w + 8\right)\left(8-3\csc^2\theta_w\right)^2}{512}\ \frac{s^2}{\left(M_Z^2-s\right)^2 + \Gamma^2M_Z^2}.
\label{coeff_Z}
\end{align}
 
\begin{table}[ht]
\centering
 \caption{The cross section values within the NLO approximation for different collision energies and renormalization scales.}
\label{tab:check_values}
\begin{tabular}{|c|c||c|c|c|c|}
\hline
\multicolumn{2}{|c||}{} &  $\sqrt{s}=0.25 M_Z$ & $\sqrt{s}=0.5 M_Z$ & $\sqrt{s}=M_z$ & $\sqrt{s}=2M_Z$\\
\hline
\hline
\multirow{2}{*}{$\mu=\sqrt{s}$} & $\sigma_{J/\psi\: \eta_c}$, fb & $3.23\cdot 10^{-2}$ & $1.23\cdot 10^{-4}$ & $2.63\cdot 10^{-5}$  & $1.91\cdot 10^{-9}$\\
& $\sigma_{J/\psi\: J/\psi}$, fb & $4.57\cdot 10^{-6}$ & $4.76\cdot 10^{-7}$ & $2.22\cdot 10^{-5}$ & $1.14\cdot 10^{-10}$ \\
\hline
\multirow{2}{*}{$\mu=10~\text{GeV}$} & $\sigma_{J/\psi\: \eta_c}$, fb & $4.88\cdot 10^{-2}$ & $2.37\cdot 10^{-4}$ & $6.84\cdot 10^{-5}$ & $6.62\cdot 10^{-9}$  \\
& $\sigma_{J/\psi\: J/\psi}$, fb & $6.87\cdot 10^{-6}$ & $9.17\cdot 10^{-7}$ & $5.84\cdot 10^{-5}$ & $4.03\cdot 10^{-10}$ \\
\hline
\end{tabular}
\end{table}

\begin{figure}[ht]
\begin{minipage}[h]{0.47\linewidth}
 \centering
\includegraphics[width = 1.1\linewidth]{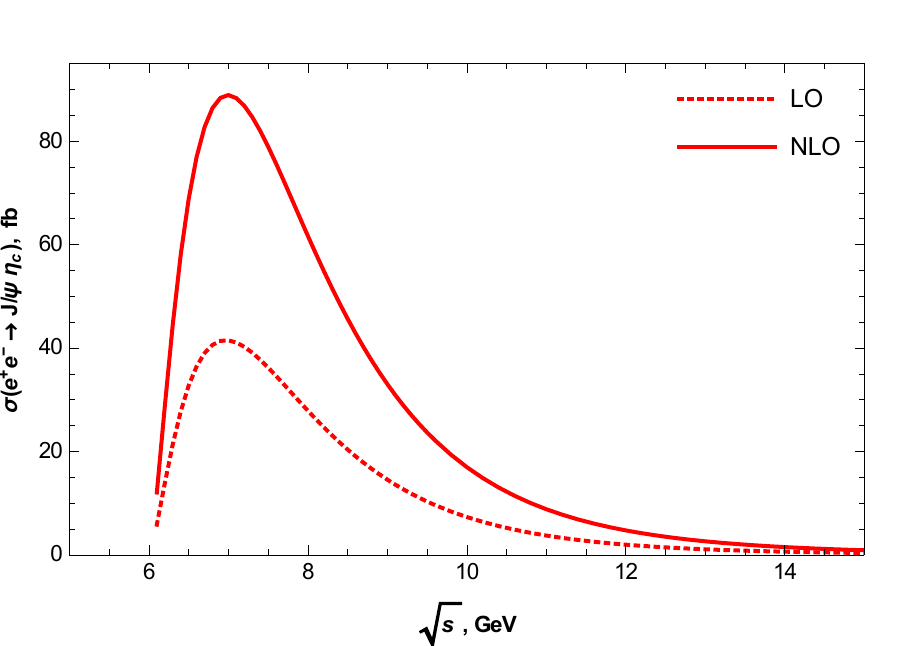}
\end{minipage}
\hfill
\begin{minipage}[h]{0.47\linewidth}
\centering
\includegraphics[width = 1.12\linewidth]{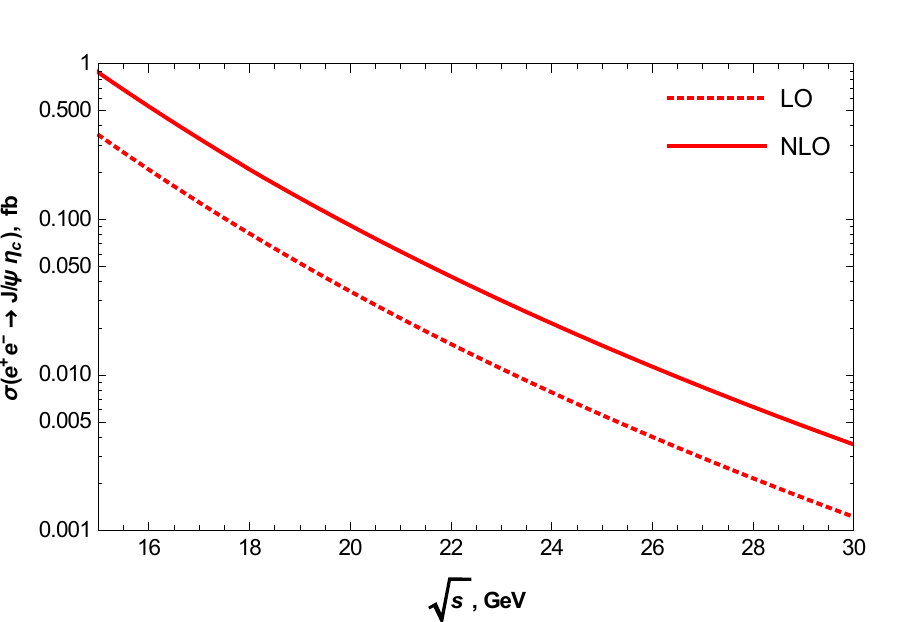}
\end{minipage}
\caption{The production cross sections for the process $e^+e^-\to\ J/\psi \: \eta_c$ calculated within the LO approximation (dashed curve) and within the NLO approximation (solid curve) as a function of interaction energy. The production cross section values are shown for the interaction energies below 30 GeV.}
\label{fig:LO_NLO}
\vfill
\begin{minipage}[h]{0.47\linewidth}
 \centering
\includegraphics[width = 1.1\linewidth]{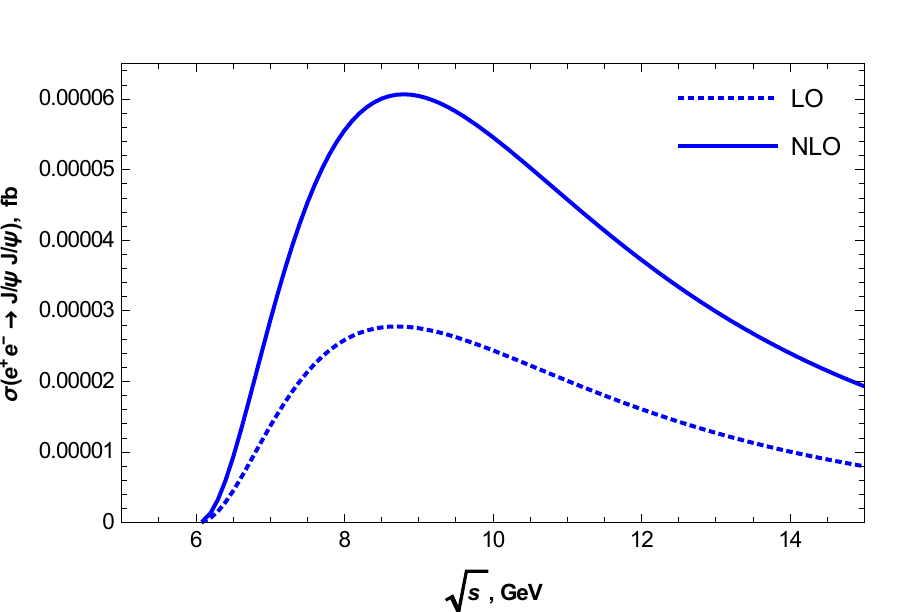}
\end{minipage}
\hfill
\begin{minipage}[h]{0.47\linewidth}
\centering
\includegraphics[width = 1.1\linewidth]{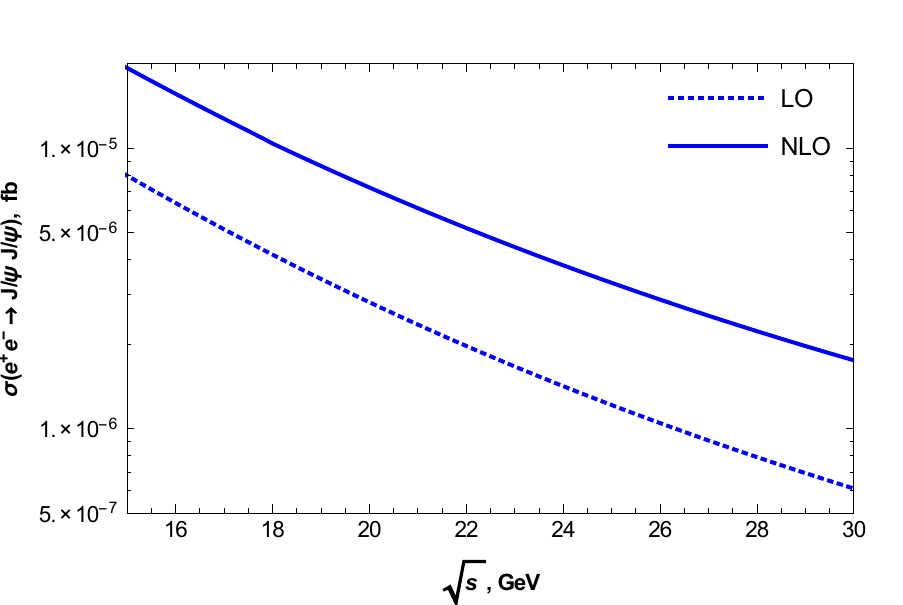}
\end{minipage}
\caption{The same as in \Cref{fig:LO_NLO}, but for  the process $e^+e^-\to\ J/\psi\: J/\psi$.}
\label{fig:LO_NLO_jpsi}
\end{figure}

\begin{figure}[ht]
\begin{minipage}[h]{0.47\linewidth}
 \centering
\includegraphics[width = 1.1\linewidth]{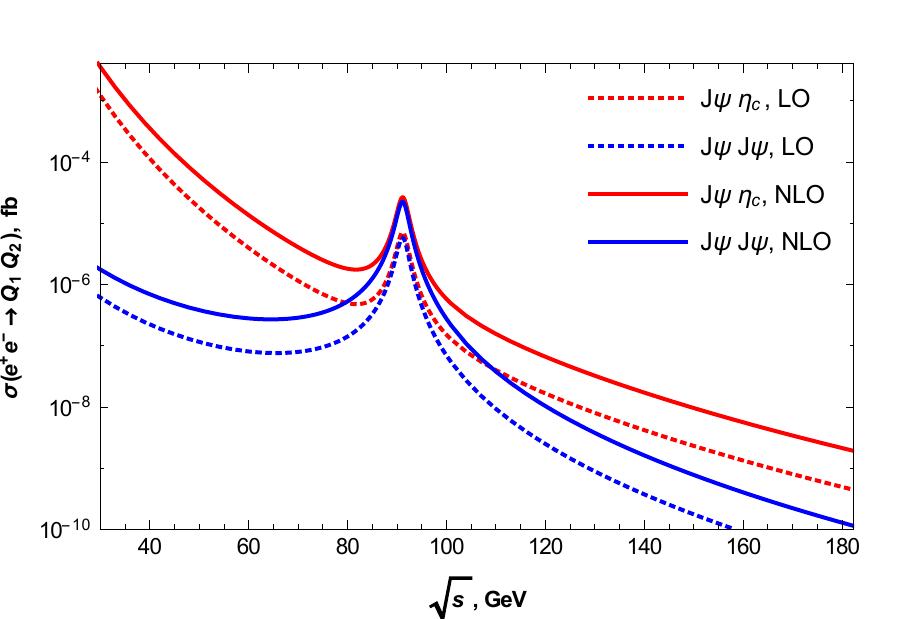}
\caption{The production cross sections for the processes $e^+e^-\to\ J/\psi \: \eta_c$ (red curves) and $e^+e^-\to\ J/\psi \: J/\psi$ (blue curves)  as a function of  interaction energy: the NLO approach (solid curves) versus the LO approach (dashed curves). The production cross section values are shown for the interaction energies above 30 GeV. }
\label{fig:NLOZpole}
\end{minipage}
\hfill
\begin{minipage}[h]{0.47\linewidth}
\centering
\includegraphics[width = 1.05\linewidth]{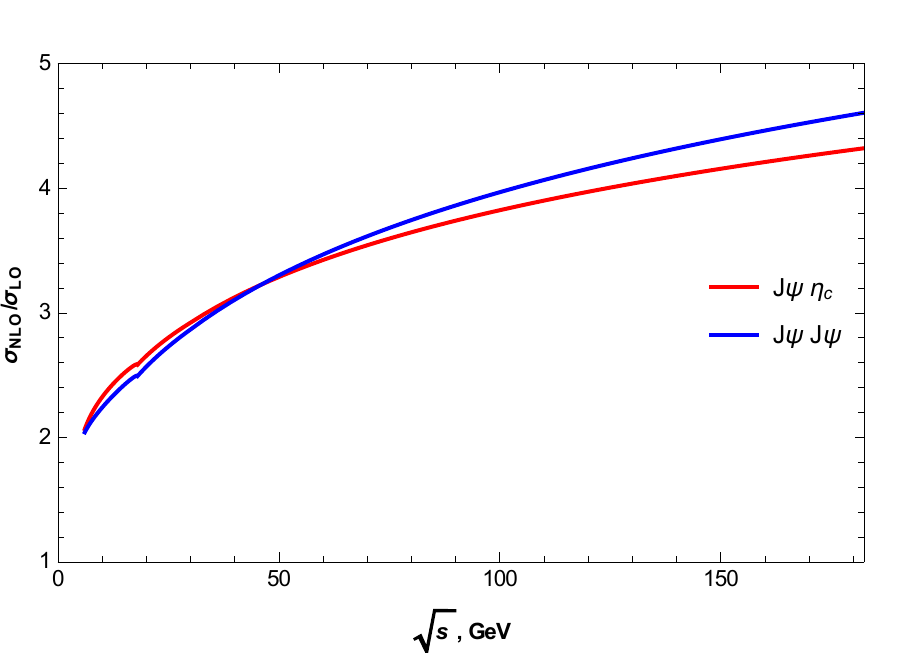}
\caption{The ratio $\sigma_{NLO}/\sigma_{LO}$ for the process $e^+e^-\to\ J/\psi \: \eta_c$ (red curve) and  $e^+e^-\to\ J/\psi \: J/\psi$ (blue curve) as a function of interaction energy.}
\label{fig:NLOtoLO}
\end{minipage}
\end{figure}

\begin{figure}[ht]
\begin{minipage}[h]{0.47\linewidth}
 \centering
\includegraphics[width = 1.1\linewidth]{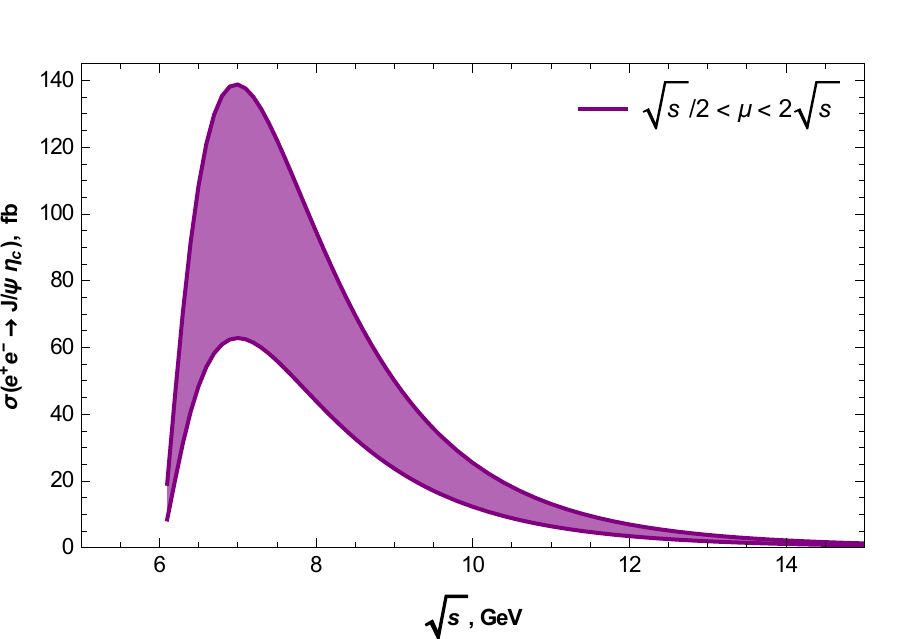}
\end{minipage}
\hfill
\begin{minipage}[h]{0.47\linewidth}
\centering
\includegraphics[width = 1.12\linewidth]{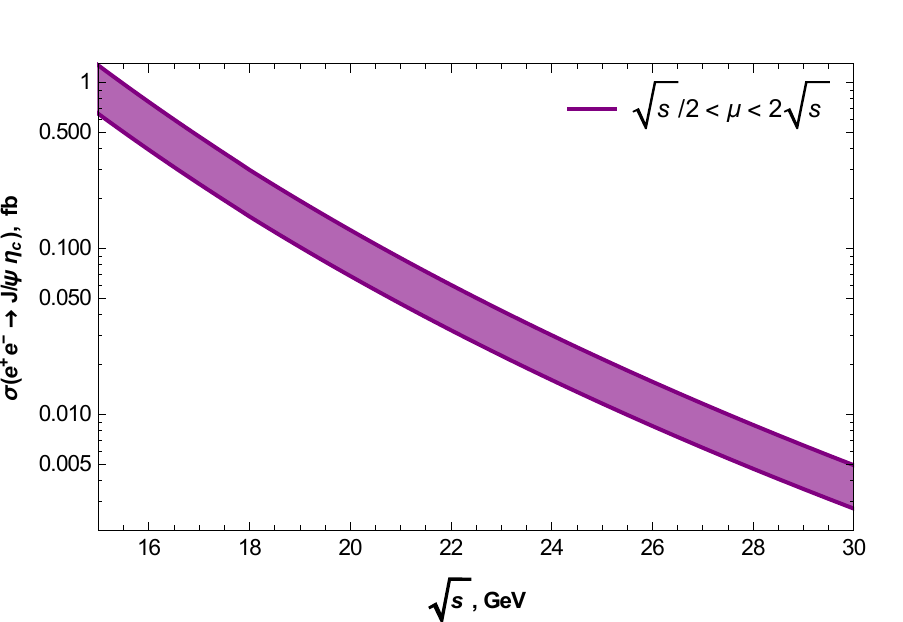}
\end{minipage}
\caption{The  dependence of NLO cross section on the interaction  energy for the process $e^+e^-\to\ J/\psi\: \eta_c$ at different scale values: $\sqrt{s} < \mu < 2\sqrt{s}$.  The production cross section values are shown for the interaction energies below 30 GeV.}
\label{fig:LO_NLODiff}
\vfill
\begin{minipage}[h]{0.47\linewidth}
 \centering
\includegraphics[width = 1.1\linewidth]{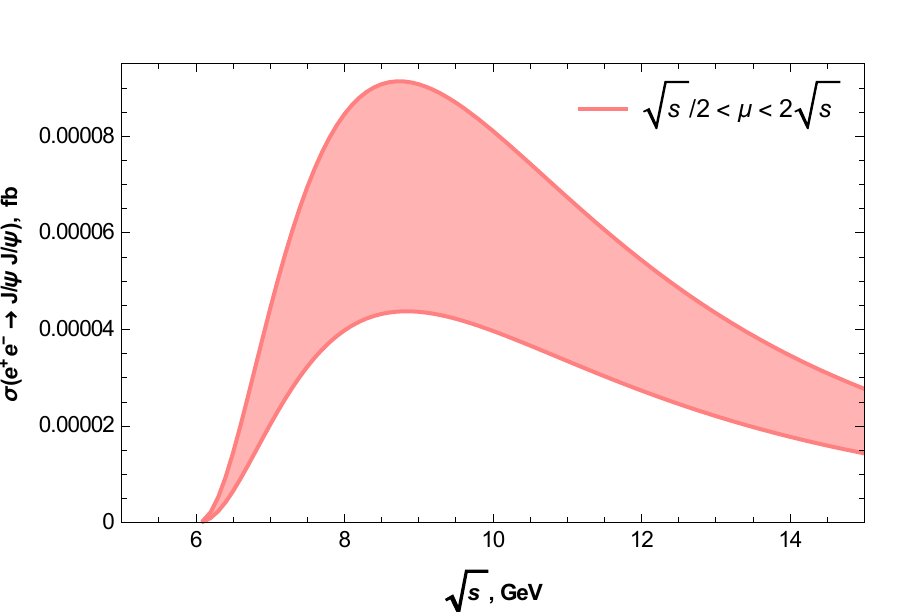}
\end{minipage}
\hfill
\begin{minipage}[h]{0.47\linewidth}
\centering
\includegraphics[width = 1.12\linewidth]{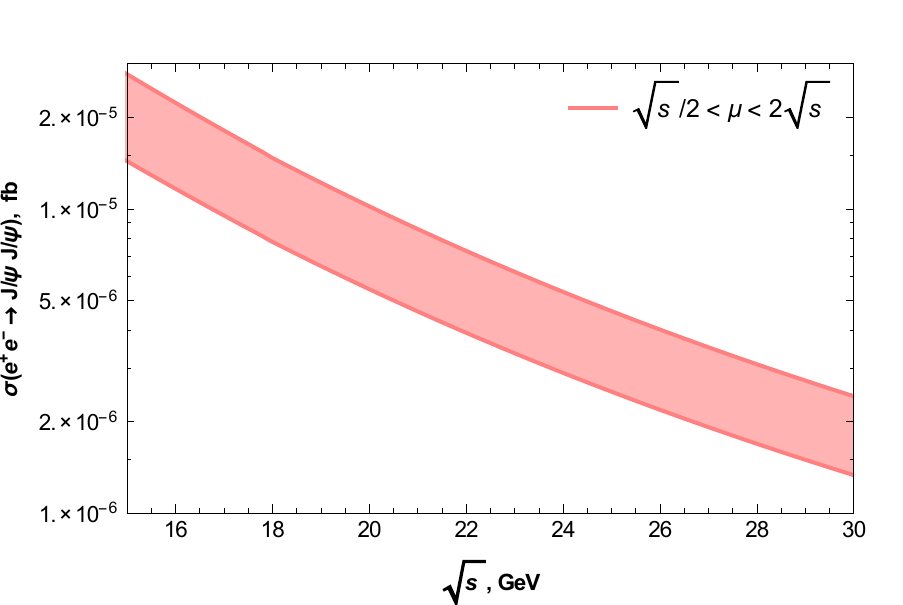}
\end{minipage}
\vfill
\caption{The same as in \Cref{fig:LO_NLODiff}, but  for the process $e^+e^-\to\ J/\psi\: J/\psi$.}
\label{fig:LO_NLODiff_jpsi}
\end{figure}

\begin{figure}[ht]
\begin{minipage}[h]{0.47\linewidth}
 \centering
\includegraphics[width = 1.1\linewidth]{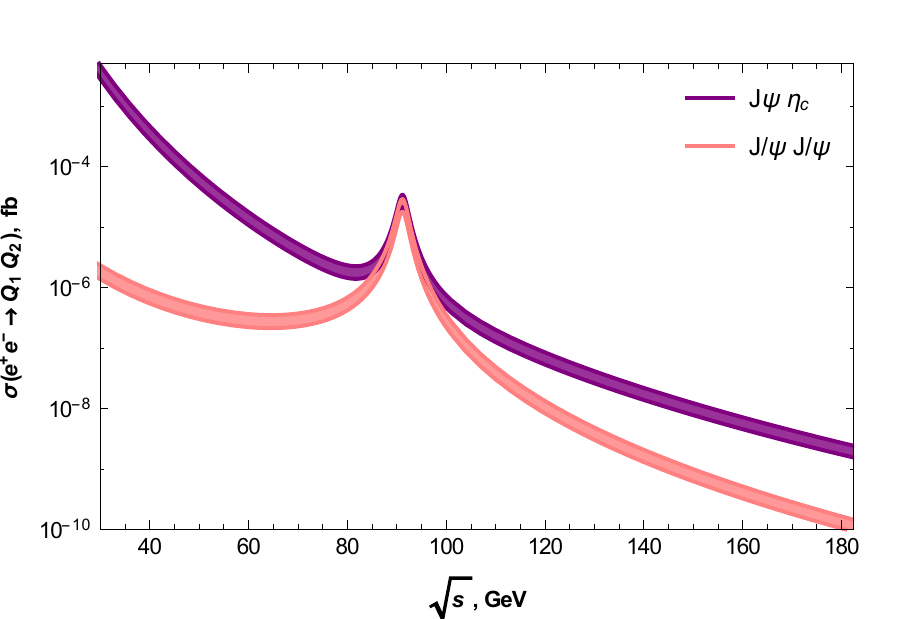}
\caption{The  dependence of NLO cross section on the interaction  energy for the processes  $e^+e^-\to\ J/\psi\: \eta_c$ (purple) and $e^+e^-\to\ J/\psi\: J/\psi$ (pink) at different scale values: $\sqrt{s} < \mu < 2\sqrt{s}$.  The production cross section values are shown for the interaction energies above 30 GeV.} 
\label{fig:NLOZpoleDiff}
\end{minipage}
\hfill
\begin{minipage}[h]{0.47\linewidth}
\centering
\includegraphics[width = 1.05\linewidth]{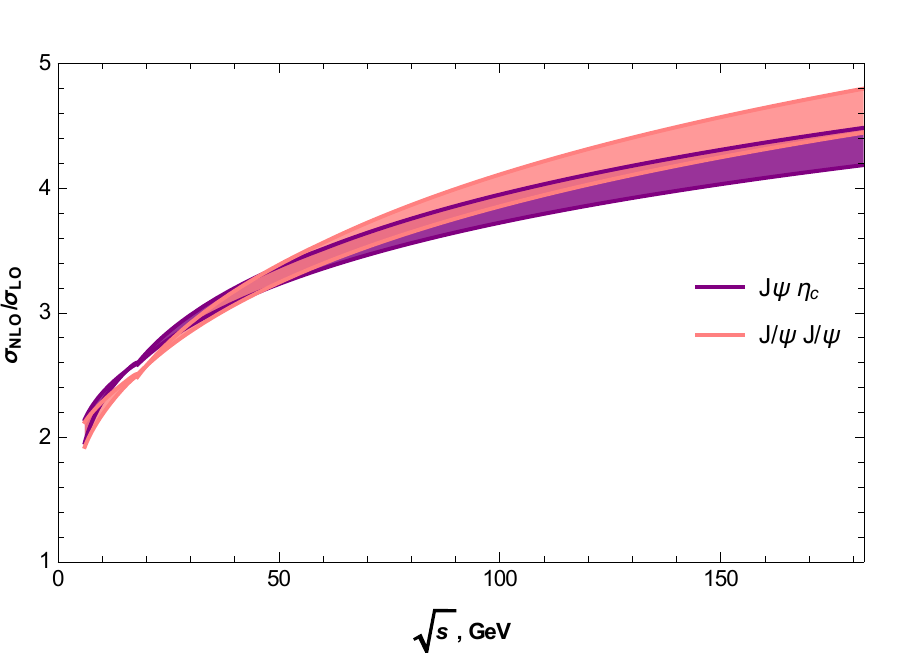}
\caption{The ratio $\sigma_{NLO}/\sigma_{LO}$ for the process $e^+e^-\to\ J/\psi \: \eta_c$ (purple) and  $e^+e^-\to\ J/\psi \: J/\psi$ (pink) as a function of interaction energy at different scale values: $\sqrt{s} < \mu < 2\sqrt{s}$.}
\label{fig:ratioDiffScales}
\end{minipage}
\end{figure}

Figures~\ref{fig:LO_NLO}--\ref{fig:NLOtoLO} demonstrate the energy dependence of the production cross sections calculated at the scale $\mu=\sqrt{s}$.
As clearly seen in the presented Figures, the NLO contributions do significantly  increase the cross section values.
To estimate the theoretical uncertainties due to the scale choice we  vary the $\mu$ value from  $\sqrt{s}/2$ to $2\sqrt{s}\ $ and present the results  in Figures~\ref{fig:LO_NLODiff}--\ref{fig:ratioDiffScales}.

Our results for $J/\psi\: \eta_c$ production at low energies do reproduce the results of earlier studies~\cite{Zhang:2005cha,Gong:2007db,Dong:2012xx,Xi-Huai:2014iaa}, what is truly essential. 

It can be seen in \Cref{fig:LO_NLO,fig:LO_NLO_jpsi}, that the cross sections of both studied processes have maximum at $\sqrt{s}\sim  7~\div 8  \mbox{ GeV}$. 

As $J/\psi\: J/\psi$ production proceeds only through the virtual $Z$,  it is expected that near the threshold such  a process will be strongly suppressed in comparison with the $J/\psi\: \eta_c$  production process. Our calculations are in agreement with this expectation: the discussed suppression is of the order $10^{-6}$ at the energies below $10~\text{GeV}$  and it lessens with the increasing energy.

Since $Z$-boson exchange dominates the area around
$Z$ pole, the cross sections  of $J/\psi\: J/\psi$ and $J/\psi\: \eta_c$ production at the corresponding energy area could be similar, in contrast to the case of the production near the threshold. 
This consideration is confirmed by our results: as seen in \Cref{fig:NLOZpole} and \Cref{tab:check_values} the cross section values are quite close to each other  near $Z$ pole.  Both LO and NLO computations of the ratio $r=\sigma_{J/\psi\: \eta_c}/\sigma_{J/\psi\: J/\psi}$   at the energy $\sqrt{s}=M_Z$  are in a good agreement with the earlier results~\cite{Likhoded:2017jmx}, where the width ratio 
$R=\Gamma\left(Z\to J/\psi\: \eta_c\right)/\Gamma\left(Z\to J/\psi\: J/\psi\right)$ was calculated within the LO accuracy:
\begin{equation}\label{ratio_in_pole}
 r_{LO} = 1.22\,, \hspace{10ex} r_{NLO} = 1.18\,, \hspace{10ex} R_{LO} = 1.20\,.
\end{equation}


Also, it is interesting to note, that as it is seen from  Figures~\ref{fig:LO_NLO_jpsi} and \ref{fig:NLOZpole}, the cross sections for the $J/\psi\: J/\psi$ production at the energies of $B$-factories and at the energies near $Z$ pole are comparable in a magnitude. 

The NLO calculations of the exclusive production of quarkonium states are known to encounter the problems related to the double logarithmic terms at high energies. In this study we confirm the result of the previous studies~\cite{Dong:2011fb,Dong:2013qw} done for the process  $e^+e^- \xrightarrow{\gamma} 
J/\psi\: \eta_c$. We have obtained the double logarithmic terms in the expansion at $\sqrt{s}>>m$ as per    
\begin{equation}
    \frac{{\cal A}^{NLO}}{{\cal A}^{LO}} \sim \alpha_S\left(c_3\ln^2s + c_2\ln s + c_1\ln\mu + c_0 \right)\,.
\end{equation}
Also we demonstrate for the first time the same behaviour both for the processes  $e^+e^- \xrightarrow{Z} 
J/\psi\: \eta_c$ and  $e^+e^- \xrightarrow{Z} 
J/\psi\: J/\psi$. The presented case is different from the $B_{c}$-pair production one. As it is shown in~\cite{Berezhnoy:2016etd}  the relative contribution of NLO QCD corrections to the $e^+e^-\to B_c^{(*)}B_c^{(*)}$ does not increase with the increasing energy. This fact requires an additional study, which we plan to perform in the future works.

Asymptotically the cross sections fall off with the increase of the energy: the LO contributions decrease as ${\cal O}\left(1/s^4\right)$ while the NLO decrease as ${\cal O}(\ln^2s/s^4)$. \Cref{fig:NLOtoLO} demonstrates that the ratio $\sigma_{NLO}/\sigma_{LO}$ increases with energy. It seems that the specific behaviour of the NLO corrections can not be compensated by the scale choice at very high energies, as explained in Ref.~\cite{Dong:2011fb,Dong:2013qw}. 
As a result, at the energies about $2M_Z$ the NLO contribution is responsible for a fivefold increase in the cross section.

\begin{figure}[ht]
\begin{minipage}[h]{0.47\linewidth}
 \centering
\includegraphics[width = 1.1\linewidth]{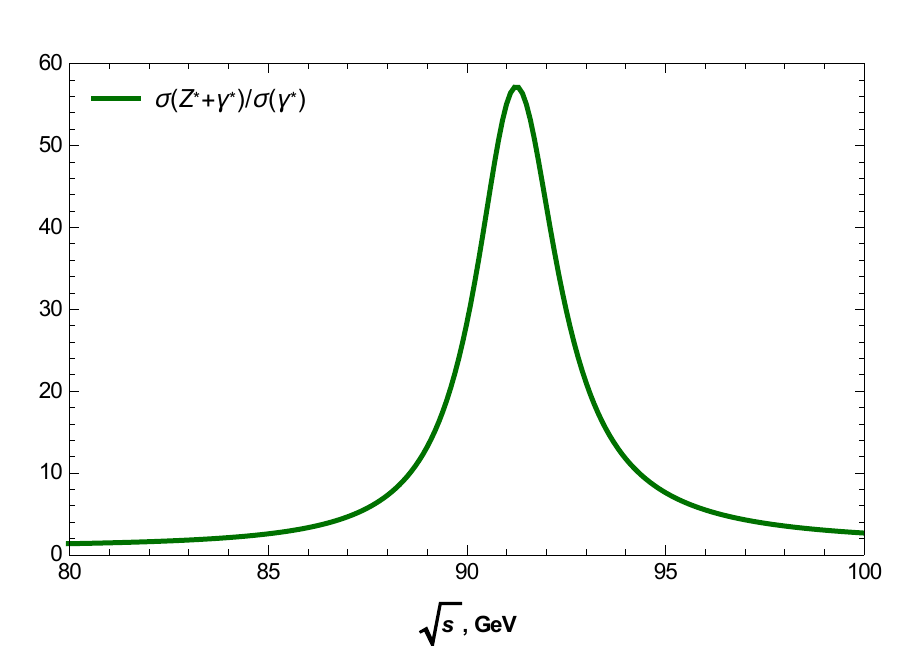}
\caption{The cross sections ratio $\sigma(\gamma^*+Z^*)/\sigma(\gamma^*)$ for the process $e^+e^-\to J/\psi\; \eta_c$  at NLO as a function of interaction energy.}
\label{fig:ZtoGamma}
\end{minipage}
\hfill
\begin{minipage}[h]{0.47\linewidth}
\centering
\includegraphics[width = 1.1\linewidth]{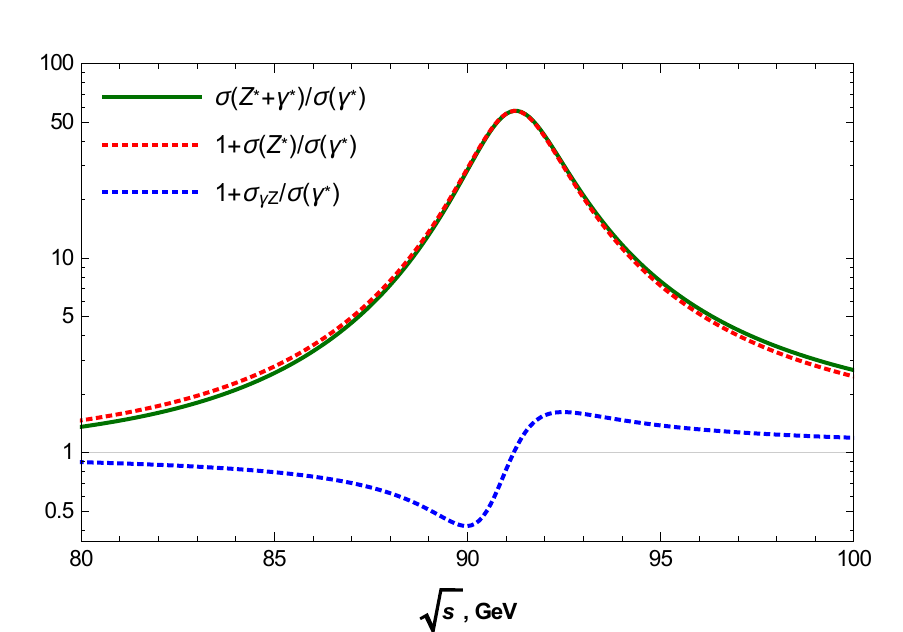}
\caption{The cross sections ratios for the  process $e^+e^-\to~J/\psi~\eta_c~$ at NLO as a function of interaction energy.}
\label{fig:ZtoGammaLog}
\end{minipage}
\end{figure}

Obviously, the  $Z$-boson exchange dominates in the $J/\psi\: \eta_c$ production around the $Z$ mass, as it is seen in \Cref{fig:ZtoGamma,fig:ZtoGammaLog}. At the $Z$ mass, 
the ratio $\sigma(Z^*+\gamma^*)/\sigma(\gamma^*)$ amounts $\approx{60}$. 
As one moves away from the  $Z$ pole, the contribution of  the  $Z$-boson  exchange diminishes in such a way that the ratio $\sigma(Z^*+\gamma^*)/\sigma(\gamma^*) > 1.1$ only in the  range $0.8 M_Z < \sqrt{s} < 2 M_Z$.

It is worth to mention that the $P$-symmetry is not violated in the considered processes, because the $V-A$ interference, which could cause such a violation,  does not  contribute either to the $J/\psi\: \eta_c$ production, or to the $J/\psi\: J/\psi$ production.

Discussing the  process $e^+e^-\xrightarrow{Z} J/\psi\: J/\psi$, it is natural to suggest a significance for the 
search of $Z\to J/\psi\: J/\psi$ decays in LHC detectors. Currently, the studies of  $Z$ decays to double quarkonia states are motivated by the CMS study~\cite{Sirunyan:2019lhe}, where  the search for Higgs and $Z$ decays to  $J/\psi$ and $\Upsilon$ pairs was performed for the first time. In this way our work complements the predictions of~\cite{Likhoded:2017jmx} and predicts that the width  $\Gamma(Z\to J/\psi\: J/\psi)$ and  $\Gamma(Z\to J/\psi\: \eta_c)$ are approximately $3.5$ times larger at the NLO approximation.

\section{CONCLUSIONS}

The cross sections of  $J/\psi\: \eta_c$ pair and $J/\psi\: J/\psi$ pair production in the $e^+e^-$ single boson annihilation are calculated within the QCD one loop approximation with the $\gamma$ exchange, the $Z$-boson exchange and the $\gamma-Z$ interference considered. It is found that the one loop QCD corrections are responsible for a significant, up to a fivefold, increase of the cross section values at all investigated energies.  It is obtained, that $\sigma_{NLO}/\sigma_{LO} \approx 3.5$ at $Z$ pole for both investigated processes. Obviously, the same enhancement by a factor of $3.5$ applies to the widths of decays $Z\to J/\psi\: J/\psi$ and $Z\to J/\psi\: \eta_c$.

The results obtained in the paper might be useful for future studies of charmonia physics at ILC and FCC colliders. Furthermore, the results are directly related to the searches of rare $Z$-boson decays into double quarkonia states in LHC detectors.

Authors would like to thank I.~Gorelov, A.~Onishchenko and A.~Luchinsky for help and constructive discussions. The work was supported by the  RFBR (grant No. 20-02-00154~A).~~I.~Belov acknowledges the support from the ``Basis'' Foundation (grant No. 20-2-2-2-1). The work of S. Poslavsky was supported by the RSCF (grant No. 20-71-00085).

\section*{References}
\bibliography{cite} 

\end{document}